%% file: 2_5D_AreaFunctions.tex
\documentclass[a4paper]{article}

\usepackage{INTERSPEECH2019}
\usepackage{graphicx}
\usepackage{array}
\usepackage{bm}
\usepackage{balance}

\newcolumntype{P}[1]{>{\centering\arraybackslash}p{#1}}

\title{An extended two-dimensional vocal tract model for fast acoustic simulation of single-axis symmetric three-dimensional tubes}
\name{Debasish Ray Mohapatra$^1$, Victor Zappi, Sidney Fels$^1$}
\address{
  $^1$
Electrical and Computer Engineering Department, University of British Columbia, Canada} 
\email{d.mohapatra@alumni.ubc.ca, victor.zappi@gmail.com, ssfels@ece.ubc.ca}

\begin{document}

\maketitle
\begin{abstract}
The simulation of two-dimensional (2D) wave propagation is an affordable computational task and its use can potentially improve time performance in vocal tracts' acoustic analysis. Several models have been designed that rely on 2D wave solvers and include 2D representations of three-dimensional (3D) vocal tract-like geometries. However, until now, only the acoustics of straight 3D tubes with circular cross-sections have been successfully replicated with this approach. Furthermore, the simulation of the resulting 2D shapes requires extremely high spatio-temporal resolutions, dramatically reducing the speed boost deriving from the usage of a 2D wave solver. In this paper, we introduce an in-progress novel vocal tract model that extends the 2D Finite-Difference Time-Domain wave solver (2.5D FDTD) by adding tube depth, derived from the area functions, to the acoustic solver. The model combines the speed of a light 2D numerical scheme with the ability to natively simulate 3D tubes that are symmetric in one dimension, hence relaxing previous resolution requirements. An implementation of the 2.5D FDTD is presented, along with evaluation of its performance in the case of static vowel modeling. The paper discusses the current features and limits of the approach, and the potential impact on computational acoustics applications.
\end{abstract}

\section{Introduction}
\input{sec_intro.tex}

\section{Methods}
\input{sec_methods.tex}

\section{Experiments and Results}
\input{sec_experiments.tex}

\vspace{-2pt}
\section{Discussion and Conclusion}
\input{sec_discussion.tex}

\vspace{-2pt}
\section{Acknowledgements}
This work is supported by the Natural Sciences and Engineering Research Council (NSERC) of Canada and Canadian Institutes for Health Research (CIHR).

\bibliographystyle{IEEEtran}

\balance

\bibliography{2_5D_AreaFunctions.bbl}


\end{document}

%% file: sec_intro.tex







The shape of the human upper vocal tract is quite intricate and the study of its acoustics proves a challenging research topic. The cavity that connects the larynx to the mouth opening forms a curved tube with varying irregular cross-sectional shape, in turn connected to several side branches, e.g., the nasal tract and the piriform fossae. Since every detail of this complex structure has an effect on voice production, the most typical approach to the analysis of vocal tract acoustics is the employment of three-dimensional (3D) simulations \cite{takemoto2010acoustic, vampola2015human, arnela2016influence}. Such simulations capitalize on the computation of pressure wave propagation within precise 3D geometrical reconstructions of real vocal tracts, and are capable of producing highly accurate results. However, the computational power required to solve wave equations in three dimensions is remarkably high, yielding to simulation times that range from several minutes to more than a day for just a few milliseconds of audio \cite{takemoto2010acoustic, arnela2014two}.

Several research groups explored two-dimensional (2D) simulations as an alternative to a precise but costly 3D approach \cite{mullen2006waveguide, speed2009characteristics, wang2012mandarin, arnela2014two}. The reason for this interest is two-fold. Compared to the 3D case, 2D wave equation solvers are characterized by much lower computational requirements and can potentially achieve real-time or quasi real-time performance \cite{mullen2007real, zappi2016towards}. Furthermore, this class of simulations appears to be capable of preserving most of the geometrical details of the vocal tract that is being analyzed. In 2D a vocal tract is represented by a flat \textit{contour}, typically extracted from the mid-sagittal slice of a 3D tube built from its area function \cite{wang2012mandarin, arnela2014two}; it can include the vocal tract's original curvature as well as 2D equivalents of its side-branches \cite{speed2009characteristics}. In other words, the only missing information regards the varying cross-sectional shape of the 3D vocal tract, which is replaced by a set of circumferences with the same area progression.

However, in a practical scenario the advantages of 2D simulations prove quite limited. As thoroughly discussed by Arnela and Guasch for the case of vowel synthesis \cite{arnela2014two}, a direct mid-sagittal representation of the tube built from the vocal tract's area function produces erroneous formant locations and bandwidths, i.e., not matching the results obtained from the 3D acoustic simulation of the same 3D shape. To fix these discrepancies, the extracted 2D contours have to undergo a non-linear deformation process that leads to a significant downsizing of the vocal tract's constrictions, in some cases up to an order of magnitude \cite{zappi2016towards}.
A direct consequence of this modification regards simulation times. To model such narrow constrictions, 2D simulations need to run with extremely high spatio-temporal resolutions, in turn increasing the computational load of the solver and producing waiting times far from real-time performance. 
Moreover, the methodology proposed by Arnela and Guasch is applicable only to the case of straight tubes, and it is still unclear how to obtain 2D contours that acoustically match curved and/or branching 3D geometries.

In this paper, we present a novel approach to vocal tract modeling, that stems from the 2D rationale \cite{zappi2016towards} and improves upon it. It capitalizes on an extended 2D Finite-Difference Time-Domain solver (2.5D FDTD), capable of simulating how pressure propagates in 3D geometries that are symmetric at least in one dimension, typically along the $z$ axis (with $x$ and $y$ being the dimensions of the starting 2D scheme). At the core of this approach there is the inclusion within the model of extra impedance terms, that derive from the tube's \textit{depth} (i.e., its continuous extension along $z$) and that are sampled in every point of the scheme; the result is a \textit{depth map}, that combined with the 2D mid-sagittal contour of the original 3D geometry allows for a fast simulation of its acoustics, with computational requirements comparable to the case of standard 2D numerical schemes.
Depth maps can be retrieved from area functions as well as from full 3D models of real vocal tracts, like Magnetic Resonance Imaging scans; furthermore, the overall 2.5D representation of the analyzed geometry leaves its original proportions intact, discarding the need for the extremely high spatio-temporal resolutions that characterizes 2D simulations. However, as discussed in the next section, the development of the 2.5D model is still in progress and at its current stage of development it can be employed only for the acoustic simulation of straight tubes with circular cross-sectional shape.

%% file: sec_methods.tex
\label{sec:methods}

\subsection{Acoustics Equation} 
\label{sec:methods:eq}
Let us consider a sound wave propagating in air, and let us call $\bm{\xi}$ the displacement vector of the medium caused by the motion of the wave and $V$ the volume of a generic element of the medium (particle). Now suppose that the medium is contained between two rigid surfaces, that extend towards infinity along both the $x$ and the $y$ axis without ever intersecting and that are one the reflected image of the other (being $x/y$ the plane of symmetry). In the described scenario, the wave moves in a space that is 3D but constrained on one axis (i.e., $z$); such a space can be described by the scalar field $D(x,y)$, which 
contains the euclidean distance between the points of the two surfaces at the same $x,y$ coordinates. In other words, $D$ describes the space in terms of its \textit{depth} along $z$.






If we choose to focus on a particle of volume $V=dxdyD$, with $dx$ and $dy$ relatively small, $\bm{\xi}$ will only have $x$ and $y$ components due to the presence of the enclosing surfaces. 
Consequently, during the passage of the sound wave, the particle's volume change can be expressed as:

\vspace{-10pt}
\begin{align}
dV = ( \frac{\partial D\xi_x}{\partial x} \ dx ) \ dy + ( \frac{\partial D\xi_y}{\partial y} \ dy )\  dx 
\label{eq:volume_change}
\end{align}
\vspace{-10pt}

\noindent 

if $\bm{\xi}$ is quite small (reasonable assumption in a realistic scenario) and $D$ is slowly varying in space. The two products in parentheses represent the first order Taylor Series approximations of the increment of $\bm{\xi}$ and $D$, on $x$ and $y$ respectively. Hence, by choosing $dx = dy = ds$, the \textit{fractional volume change} can be written as:

\vspace{-10pt}
\begin{align}
\frac{dV}{V} \ = \ \frac{1}{D} \ ( \ \frac{\partial D\xi_x}{\partial x} \ + \ \frac{\partial D\xi_y}{\partial y} \ )
\label{eq:frac_vol}
\end{align}
\vspace{-10pt}

\noindent 
If we describe the motion of the particle via Newton's equation and we combine it with the equation of state $p \ = -K\frac{dV}{V}$ (where $K$ is the bulk modulus of air), it is possible to prove the following relationship between the displacement and the fractional volume change just obtained:

\vspace{-10pt}
\begin{align}
\frac{\partial^2 \bm{\xi}}{\partial t^2} \ = \ \frac{c^2}{D} \bm{\nabla} ( \ \frac{\partial D\xi_x}{\partial x} \ + \ \frac{\partial D\xi_y}{\partial y} \ )
\label{eq:wave_disp}
\end{align}
\vspace{-10pt}

\noindent with $c$ defined as the speed of sound in air. Finally, by following Fletcher and Rossing \cite{fletcher2012physics} and considering $D$ independent of time, we can obtain from Equation \ref{eq:wave_disp} the following acoustic wave equation:

\vspace{-10pt}
\begin{align}
\frac{\partial^2 p}{\partial t^2} \ = \ \frac{c^2}{D}  (  \ \frac{\partial}{\partial x} (D \frac{\partial p}{\partial x}) \ + \ \frac{\partial}{\partial y} (D \frac{\partial p}{\partial y}) \ )
\label{eq:wave_extended}
\end{align}
\vspace{-10pt}

\noindent where $p$ is sound pressure. The equation describes air wave motion in a 3D space constrained in one dimension; given the similarity with the standard 2D acoustic wave equation, this 3D space can be considered as an \textit{extended} 2D (2.5D) space. By symmetrically constraining the 2.5D space along the $y$ dimension too (hence turning $D$ into a circular area), and by assuming plane-wave propagation only (due to large distance from the source), Equation \ref{eq:wave_extended} reduces to Webster's horn equation \cite{ungeheuer2013elemente}.

\subsection{2.5D FDTD Wave Solver}
\label{sec:methods:solver}
The procedure followed for the design of the 2.5D wave solver shares many similarities with the case of the 2D vocal tract model described in \cite{zappi2016towards}.
We decomposed the 2.5D acoustic wave equation into extended versions of the 2D equation of continuity and 2D equation of motion:

\vspace{-10pt}
\begin{align}
\frac{\partial p}{\partial t} \ = \ -  \frac{\rho c^2}{D} (  \ \frac{\partial D v_x }{\partial x}  \ + \ \frac{\partial D v_y }{\partial y} \ ) 
\label{eq:wave_cont}\\
\beta\frac{\partial\bm{v}}{\partial t} \ + \ \left(1-\beta\right)\bm{v} \ = \ -\beta^{2}\frac{\bm{\nabla} p}{\rho} \ + \ \left(1-\beta\right)\bm{v}_{b}
\label{eq:wave_mot}
\end{align}
\vspace{-10pt}

\noindent where $\bm{v}$ is the 2D acoustic particle velocity. Equation \ref{eq:wave_mot} does not include any explicit dependence on the depth of the considered space (via $D$). This is due to the fact that the equation of motion derives from Newton's second law, whose 2D and 2.5D forms coincide. However, similarly to what proposed by Allen and Raghuvanshi \cite{allen2015aerophones}, we augmented the obtained standard 2D equation with the scalar field $\beta(x, y, t)$. By varying between 1 and 0, this term allows for the transition between the momentum equation and the enforcement of a prescribed velocity $\bm{v_b}$. Through this mechanism it is possible to enforce boundary conditions (more details in the next paragraph) and to simulate dynamic geometries \cite{zappi2016towards}.

We applied a standard 2D Yee scheme, where each grid point consists of a squared 2D cell \cite{allen2015aerophones, yee1966numerical}. Per each time step $n$, pressure values $p^{(n)}$ are sampled across the whole domain at the center of the cells, while the $v_x^{(n)}$ and $v_y^{(n)}$ components of the velocity vectors are sampled on the edges shared with the right and top neighbor cell respectively. Hence, in a generic cell at discrete coordinates $(\hat{x}, \ \hat{y})$, $p$ will be sampled at $(\hat{x}, \ \hat{y})$, while $v_x$ at $(\hat{x}+1/2, \ \hat{y})$ and $v_y$ at $(\hat{x}, \ \hat{y}+1/2)$. This leads to the following discrete update rules of Equations \ref{eq:wave_cont} and \ref{eq:wave_mot}, where we denote with $\bm{\widetilde{\nabla}}$ the standard discrete spatial derivatives as performed in 2D FDTD:

\vspace{-10pt}
\begin{align}
& p^{(n+1)} \ = \ \frac{ \bar{D} p^{(n)}  - \rho c^2 \Delta  t \ \bm{\widetilde{\nabla}} \cdot \bm{V}^{(n)} } { \bar{D}} 
\label{eq:fdtd_cont}\\
&\bm{v}^{(n+1)} \ = \ \frac{ \beta \bm{v}^{(n)}  - \beta^2 \Delta  t \ \widetilde{\nabla} p^{(n+1)} / \rho \ + \  \Delta t(1-\beta) \bm{v_b}} { \beta + \Delta t(1-\beta)} 
\label{eq:fdtd_mot}
\end{align}
\vspace{-17pt}

with:

\vspace{-17pt}
\begin{flalign}
& \bm{V} \ = \ ( \ D_{(x)} v_x,  \ D_{(y)} v_y \ )
\label{eq:extended_vel}
\end{flalign}
\vspace{-10pt}

\noindent The terms $\bar{D}$, $D_{(x)}$ and $D_{(y)}$ are the components of the depth map of the 2.5D space and they effectively act as extra impedance terms in two dimensions. Their relation with the field $D$ will be explained in detail in the next subsection. 

In line with the work of Takemoto and Mokthari \cite{takemoto2010acoustic}, the vocal tract's walls are simulated by adapting the local reactive boundary approach originally proposed by Yokota et al. \cite{yokota2002visualization}. This is done by means of setting $\beta = 0$ and $\bm{v_b} = \rho c \ \mu \ p_w \bm{n}$, with $p_w$ equal to the pressure value sampled from the cell in front of the wall, $\bm{n}$ the unit vector normal to it and directed towards the wall itself and $\mu$ the boundary admittance. Similarly, arbitrary excitation can be injected into the domain via boundary cells (i.e, $\beta = 0$), by setting $\bm{v_b}$ equal to the output velocity of a \textit{glottal model}; this can be coupled with the vocal tract by feeding back the pressure value of the cell in front of the excitation velocity stream. Alike its 2D version \cite{zappi2016towards}, the 2.5D vocal tract model allows for the coupling with both 1D \cite{ishizaka1972synthesis, story1995voice} and 2D glottal models \cite{vasudevan2017fast}.

\subsection{Depth Map}
\label{sec:methods:depth}
The depth map is a discretized version of the continuous field $D$, processed according to the described 2D scheme. Since every individual FDTD cell holds three acoustic parameters sampled in three different locations (center, right and top), the discretization of $D$ will likewise produce three depth values per each cell, namely $\bar{D}$, $D_{(x)}$ and $D_{(y)}$. 
As a result, in the FDTD grid the pressure values are aligned with $\bar{D}$ values, while the velocity vector components are aligned with $D_{(x)}$ and $D_{(y)}$ values.

\begin{figure}[!t]
	\centering
	\includegraphics[clip, trim= 0cm 0cm 0cm 0cm, width=\columnwidth]{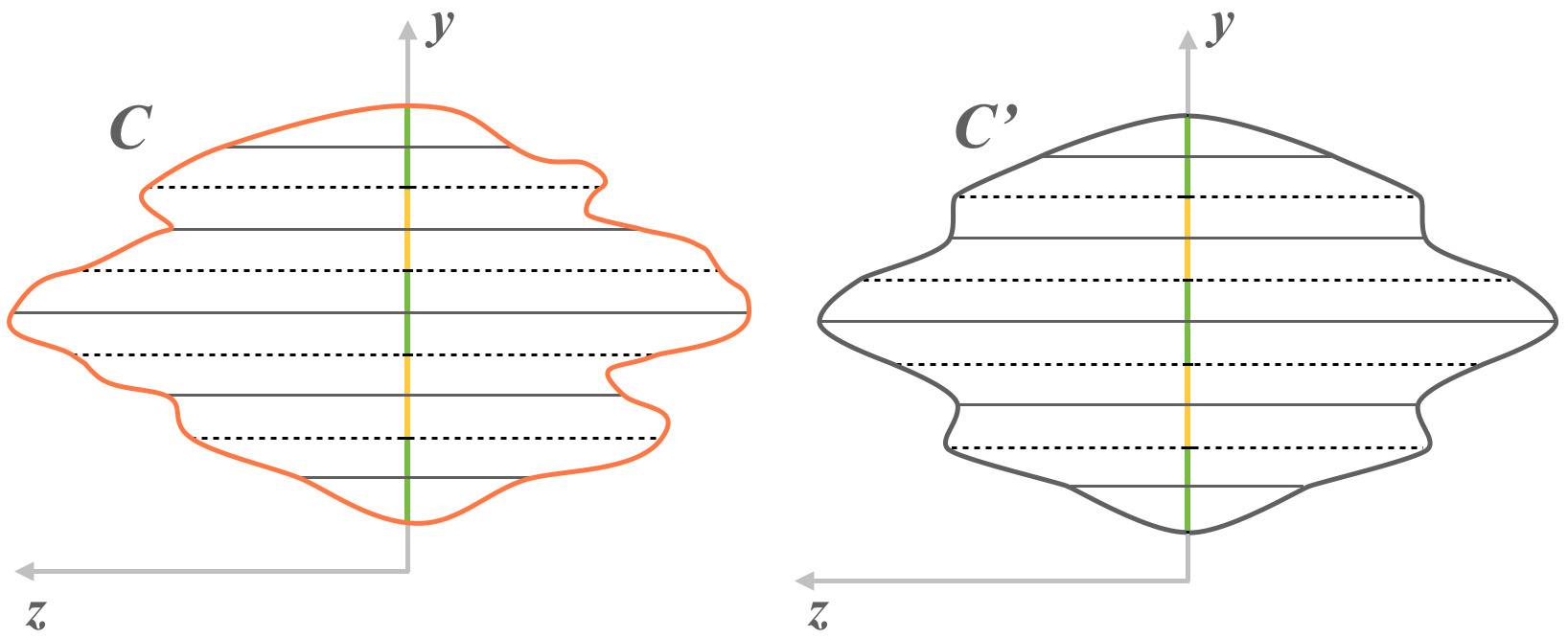}
	\vspace{-0.5 cm}
	\caption{on the left, depth map extraction ($D_{(x)}$ solid line, $D_{(y)}$ dashed line) from a slice $C$ of an irregular tube, at coordinate $x=\hat{x}$ (for simplicity we assume the same shape at $\hat{x}+1/2$). On the right, equivalent single-axis symmetric cross-section $C'$, as represented in the 2.5D FDTD.}
	\label{fig:slice}
\vspace{-15 pt}
\end{figure}

In a practical scenario, the field $D$ can be replaced by the 3D model of the geometry whose acoustics is being analyzed.
Even if typically composed of discrete elements (e.g., vertices), the model can be treated as a continuous geometry via interpolation. However, $D$ always represents a symmetric domain; this is not necessarily true for a 3D model, hence additional processing is required to derive the depth map from it. In the case of a generic 3D tube, the depth map is computed as follows:

\vspace{-2pt}
\begin{enumerate}
	\item a 2D contour is extracted as the intersection between the tube and its mid-sagittal plane (i.e., 2D vocal tract geometry).
          The contour itself represents the vocal tract's walls, while the mid-sagittal plane becomes the $x/y$ plane of the 2.5D space;

    	\item in every wall cell, $\bar{D}$, $D_{(x)}$ and $D_{(y)}$ are set to zero;

    	\item in every cell outside of the the tube's walls, $\bar{D}$, $D_{(x)}$ and $D_{(y)}$ are assigned a fixed \textit{open space depth};
    
    	\item for every cell with discrete coordinates $(\hat{x}, \ \hat{y})$ inside the tube's walls, the two lines perpendicular to the $x/y$ plane and passing by $(\hat{x}+1/2, \ \hat{y})$ and $(\hat{x}, \ \hat{y}+1/2)$ are intersected with the model (Figure \ref{fig:slice}). The length of the two resulting segments is assigned to $D_{(x)}(\hat{x}, \ \hat{y})$ and $D_{(y)}(\hat{x}, \ \hat{y})$ respectively;

	\item for every cell with discrete coordinates $(\hat{x}, \ \hat{y})$ inside the tube's walls, $D_{(x)}(\hat{x}, \ \hat{y})$ and $D_{(y)}(\hat{x}, \ \hat{y})$ are interpolated as $( \ D_{(x)}(\hat{x}, \ \hat{y}) + D_{x}(\hat{x}+1, \ \hat{y}) \ ) / 2$ and $( \ D_{(y)}(\hat{x}, \ \hat{y}) + D_{(y)}(\hat{x}, \ \hat{y}+1) \ ) / 2$ respectively;

    \item  for every cell with discrete coordinates $(\hat{x}, \ \hat{y})$ inside the tube's walls, the depth $\bar{D}(\hat{x}, \ \hat{y})$ is obtained as $( \ D_{(x)}(\hat{x}, \ \hat{y}) + D_{(x)}(\hat{x}-1, \ \hat{y}) + D_{(y)}(\hat{x}, \ \hat{y}) + D_{(y)}(\hat{x}, \ \hat{y}-1) \ ) / 4$;

    \item a \textit{minimum depth threshold} is defined and applied in every cell (depth values are \textit{clamped}). A typical threshold is at least an order of magnitude lower than the smallest non-zero depth value found across the domain.
\end{enumerate}

The result of this process is a single-axis symmetric equivalent of the original 3D model. The cross-sections of such a 2.5D shape have same areas and widths (i.e., extension along $y$) as the 3D model's sections; moreover, their shapes can be irregular, but always symmetric along $z$, as showed in Figure \ref{fig:slice}.

The interpolations at steps 5 and 6 are introduced to satisfy the assumption that $D$ is slowly varying in space, required to obtain Equation \ref{eq:volume_change} (Section \ref{sec:methods:eq}). This enhances the overall stability of the solver too.

\vspace{-5 pt}

\subsection{Model Applicability} 
\label{sec:methods:app}
The development of the 2.5D vocal tract model is still in progress, yet some of its features can be analyzed already.

The use of the depth map imposes to slightly change the cross-sectional shapes of the original geometry by means of single-axis symmetry. As a result, the acoustics of the 3D and of the 2.5D vocal tracts will be somewhat different in the portion of the spectrum that is beyond 5 kHz \cite{arnela2016influence}. 
However, the 2.5D representation is capable of preserving curvature on the $x/y$ plane, area progression as well as  several details of the varying irregular cross-sectional shapes of real vocal tract geometries. These are features beyond the capabilities of 1D and 2D models, and whose combined acoustics effects are yet to be studied.

At the current stage of development, using the 2.5D model to analyze the acoustics of realistic vocal tracts is impractical though. The model capitalizes on the solution of a lossless system (Equations \ref{eq:wave_cont}, \ref{eq:wave_mot}). The solver adds wall impedance by means of the 2D walls boundary condition (Section \ref{sec:methods:solver}), but the losses happening on the surface described by the depth map (i.e., 2.5D boundaries) are largely ignored, as the extra depth-related impedance terms $\bar{D}$, $D_{(x)}$ and $D_{(y)}$ model only the effects of the surface's spatial derivative\footnote{This scenario is analogous to the case of Webster's horn equation, where the absolute size of the modeled tube does not affect propagation.}. In other words, there is no consistent way of matching the non-planar modes of a generic 3D shape.

This issue can be easily overcome for the modeling of 3D tubes with circular cross-sections, which can also include bending. By extending the calculation of modes in plates proposed by Fletcher and Rossing \cite{fletcher2012physics}, it is possible to prove that non-planar modes in each section of a lossless 2.5D tube are identical to the ones found in its 2D mid-sagittal cross-section. As a result, we can apply the methodology proposed by Arnela and Guasch \cite{arnela2014two} to match the first non-planar mode of a circular cross-section, by scaling the radii of every section of the tube by the constant factor $\frac{0.5\pi}{1.84}$ and by modifying the depth map accordingly.


%% file: sec_experiments.tex

\subsection{Model Validation}
A comparative study was carried out to validate the model's precision, by using as reference a high-accuracy 3D Finite Element Method (FEM) \cite{arnela2013finite}. 
To this end, we computed the transfer functions of a set of vocal tract-like geometries, we extracted the corresponding formants' positions and compared them with the results obtained with the 3D FEM. 
The study was carried out for the following static vowels: /a/, /i/ and /u/.
We chose to construct the 3D tubes from Story's area functions \cite{story2008comparison}. This dataset defines a standard in vocal tracts' acoustic analysis and is ideal to work with circular cross-sections (see Section \ref{sec:methods:app}).

As further validation, we compared the model's time performance with two different versions of the same 2D FDTD:  a standard serial implementation (2DS), and a highly optimized parallel implementation (2DP). Both the models were timed while computing the transfer function of Story's /u/.
The target of such comparisons was to estimate the extra computational cost introduced by the depth terms, as well as the time boost made possible by a parallel pipeline.


\begin{figure}[!t]
\includegraphics[width=\columnwidth]{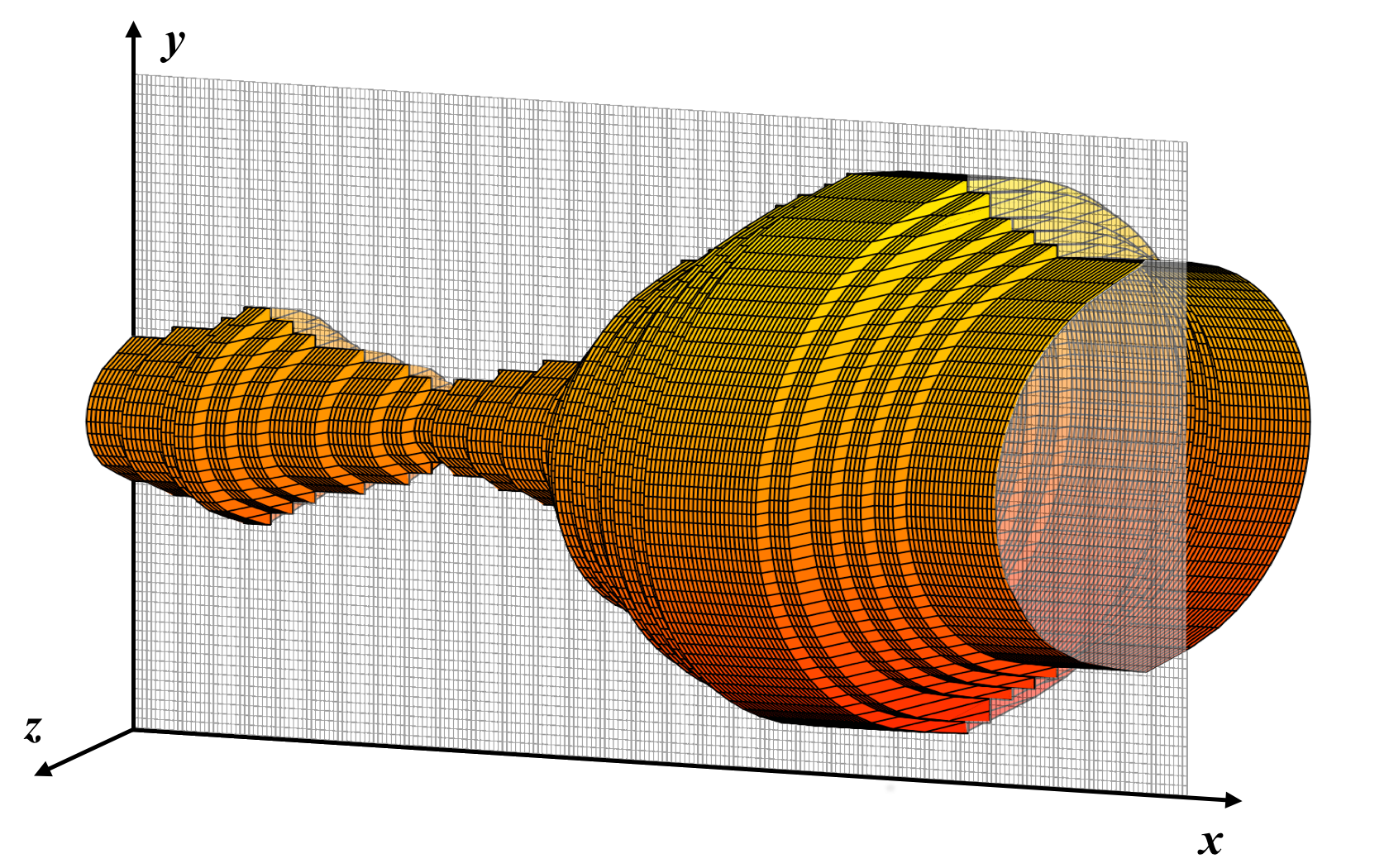}
\vspace{-20pt}
\caption{2.5D tube discretization (mid-sagittal contour and depth map) for vowel /a/. }
\label{fig:2_5Dgeometry}
\vspace{-17pt}
\end{figure}

\subsection{Experimental Setup}
The 2.5D FDTD acoustic wave solver was implemented in MATLAB. 
The resolution of the FDTD grid was set to $\Delta s = 0.74$ mm; this is the largest value that preserves the geometrical details of the tubes and was determined empirically. Any resolution above this threshold produces less accurate results. 
Figure \ref{fig:2_5Dgeometry} illustrates the resulting 2.5D tube geometry for vowel /a/.
We set the size of the grid to $270 \times 45$ cells, to obtain a domain that could fit any of the three tubes.
The temporal resolution $\Delta t$ of the simulation is restricted by the Courant-Friedrichs-Lewy condition in 2D, $\Delta t \leq \Delta s/\sqrt{2}c$, where $c$ is the speed of sound ($c=350$ m/s within the vocal tract).
We therefore set $\Delta t=1.51 \times 10^{-6}$ s, equal to a simulation rate of 661,500 Hz.

Air density was set to $\rho=1.14$ kg/m\textsuperscript{2}s and all wall cells were assigned a standard boundary admittance $\mu=0.005$.
Since the 2.5D FDTD does not yet include a way to precisely model the radiation losses, we implemented an open-end termination at the mouth opening by imposing a Dirichlet boundary condition in the same fashion of what done in \cite{arnela2014two}. 
Following the procedure described in \cite{zappi2016towards}, the impulse responses were obtained by exciting the model with a band-passed velocity pulse injected from the glottal end; the pressure variation was recorded via a microphone placed 3 mm inside the mouth opening, keeping track of computational times.
Per each vocal tract, the MATLAB solver serially iterated across the full grid to simulate a total of 50 ms of audio. The application ran on a workstation equipped with an Intel Core i7-8700K processor.


The same parameters were maintained also for the 2DS and the 2DP, but all the depth-related impedance terms were removed from their solvers\footnote{These settings produce incorrect 2D formants. Yet, the 2D models were used for time comparisons only and their output was ignored.}. Furthermore, while the 2DS was implemented in MATLAB, the 2DP was implemented as a shader as described in \cite{zappi2016towards} and ran on a Nvidia GTX 1080 graphics card.

\vspace{-5pt}

\subsection{Results}
We applied the Fast Fourier Transform to the impulse responses to obtain the vocal tracts' transfer functions for the three vowels. The positions of the first 8 formants were extracted and compared with the results of the 3D FEM approach. The exact formants' positions for the 3D FEM can be found in Arnela's PhD dissertation \cite{thesis2015Arnela}. Table \ref{tab: formant_error} shows the formants' positional differences and the percentage error between the 2D and the 3D simulation results for /a/, /i/ and /u/ respectively. For each vowel, the 2.5D run-time for producing 50 ms of audio was 15 minutes and 21 seconds;
for the 2DS it was 14 minutes and 41 seconds, while for the 2DP 1.4 seconds.

\begin{table}[!t]
    \begin{tabular}{|P{0.17\linewidth}|P{0.20\linewidth}|P{0.20\linewidth}|P{0.20\linewidth}|}
    \hline
    \textbf{Formants} & \textbf{/a/} & \textbf{/i/} & \textbf{/u/} \\ \hline 
    F1 & \begin{tabular}[c]{@{}c@{}}4Hz\\ 0.57\%\end{tabular} & \begin{tabular}[c]{@{}c@{}}-3Hz\\ -1.14\%\end{tabular} & \begin{tabular}[c]{@{}c@{}}1Hz\\ 0.38\%\end{tabular} \\ \hline
    F2 & \begin{tabular}[c]{@{}c@{}}-28Hz\\ -2.62\%\end{tabular} & \begin{tabular}[c]{@{}c@{}}49Hz\\ 2.32\%\end{tabular} & \begin{tabular}[c]{@{}c@{}}-37Hz\\ -4.88\%\end{tabular} \\ \hline
    F3 & \begin{tabular}[c]{@{}c@{}}-11Hz\\ -0.36\%\end{tabular} & \begin{tabular}[c]{@{}c@{}}50Hz\\ 1.66\%\end{tabular} & \begin{tabular}[c]{@{}c@{}}36Hz\\ 1.59\%\end{tabular} \\ \hline
    F4 & \begin{tabular}[c]{@{}c@{}}56Hz\\ 1.37\%\end{tabular} & \begin{tabular}[c]{@{}c@{}}82Hz\\ 1.98\%\end{tabular} & \begin{tabular}[c]{@{}c@{}}17Hz\\ 0.47\%\end{tabular} \\ \hline
    F5 & \begin{tabular}[c]{@{}c@{}}-59Hz\\ -1.17\%\end{tabular} & \begin{tabular}[c]{@{}c@{}}1Hz\\ 0.01\%\end{tabular} & \begin{tabular}[c]{@{}c@{}}107Hz\\ 2.56\%\end{tabular} \\ \hline
    F6 & \begin{tabular}[c]{@{}c@{}}-23Hz\\ -0.40\%\end{tabular} & \begin{tabular}[c]{@{}c@{}}-13Hz\\ -0.22\%\end{tabular} & \begin{tabular}[c]{@{}c@{}}-102Hz\\ -2.01\%\end{tabular} \\ \hline
    F7 & \begin{tabular}[c]{@{}c@{}}-5Hz\\ -0.07\%\end{tabular} & \begin{tabular}[c]{@{}c@{}}-95Hz\\ -1.44\%\end{tabular} & \begin{tabular}[c]{@{}c@{}}-46Hz\\ -0.75\%\end{tabular} \\ \hline
    F8 & \begin{tabular}[c]{@{}c@{}}25Hz\\ 0.32\%\end{tabular} & \begin{tabular}[c]{@{}c@{}}35Hz\\ 0.45\%\end{tabular} & \begin{tabular}[c]{@{}c@{}}26Hz\\ 0.39\%\end{tabular} \\ \hline
    \end{tabular}
    \vspace{1mm}
    \caption{positional errors of the first 8 formants in 2.5D, computed for vowel /a/, /i/, and /u/ with respect to 3D FEM values.} \label{tab: formant_error}
\vspace{-30pt}
\end{table}

%% file: sec_discussion.tex
The acoustic analysis of vowels' area functions is a standard validation methodology that allows to test the fundamental properties of a wave propagation model. In the case of the 2.5 FDTD, all the three vowels are characterized by positional errors that tend to stay below 2\%, and in several cases do not go above 1\%. 
To achieve comparable results, a standard 2D FDTD requires a spatial resolution that is almost three times higher than the value used in this test \cite{ zappi2016towards}. 
Such a level of accuracy suggests that 2.5D equations are a valid means to model how waves travel in a constrained space, and that the proposed solver is correct. Furthermore, the affordable spatial resolution remarkably decreases the computational load of the simulation compared to 2D models' massive grids \cite{arnela2014two, wang2012mandarin, zappi2016towards}.

The current serial MATLAB implementation of the 2.5D FDTD proves to be extremely lightweight, even when compared with the performance of the equally-sized 2DS. The time difference between the two systems is 20 s (2\%); this value is quite 
small and derives from the fact that the instruction sets of the 2.5D and 2DS solvers differ by four multiplications only (see \cite{zappi2016towards}). Given the similar computational load of the two systems, it is safe to assume that an optimized parallel implementation of the 2.5D
running on a modern graphics card will reach quasi-real-time performance, close to what achieved with the 2DP. 

These results gain particular interest in the context of the future development of the 2.5D vocal tract model. 
We are currently working on the inclusion of free-radiation effects (via 2.5D perfect matching layers) and on the validation of the modeling of the vocal tract's curvature. The inclusion of these features is relatively straightforward and will make the system capable of simulating geometrical and acoustic details still unavailable in one-dimensional and 2D models. The following step will consist of the introduction in the acoustic equations of extra loss terms, to model 2.5D boundaries. By doing so, it will be possible to simulate the effects of varying irregular cross-sections, thus fully exploring the potential of the 2.5D model.  